\documentclass[a4paper,12pt]{article}

\usepackage{comment}
\usepackage{fancyvrb}
\usepackage{graphicx}
\usepackage{hyperref}
\usepackage[ragged]{footmisc}
\usepackage{ragged2e}
\usepackage[section]{placeins}

\begin{document}

\title{The curse of variety in computing, and what can be done about it}

\author{J Gerard Wolff\footnote{Dr Gerry Wolff, BA (Cantab), PhD (Wales), CEng, MBCS, MIEEE; CognitionResearch.org, Menai Bridge, UK; \href{mailto:jgw@cognitionresearch.org}{jgw@cognitionresearch.org}; +44 (0) 1248 712962; +44 (0) 7746 290775; {\em Skype}: gerry.wolff; {\em Web}: \href{http://www.cognitionresearch.org}{www.cognitionresearch.org}.}}

\maketitle

\section{Key insights}

\begin{itemize}

    \item Excess freedom in how computers are used creates problems that include: bit rot, problems with big data, problems in the creation and debugging of software, and problems with cyber security.

    \item To tame excess freedom, ``tough love'' is needed in the form of a {\em universal framework for the representation and processing of diverse kinds of knowledge} (UFK).

    \item The {\em SP machine}, based on the {\em SP theory of intelligence}, has the potential to provide that framework and to help solve the problems above.

    \item There is potential to reduce the near-4000 different kinds of computer file to one, and to reduce the hundreds of different computer languages to one.

\end{itemize}

\noindent\rule{\textwidth}{0.4pt}
\vspace{3mm}

Imagine that, instead of two or three international standards for electrical plugs and sockets, there were thousands. Imagine that every town and village had its own standard, that each standard was incompatible with any other standard, and that each electrical appliance might come fitted with a plug conforming to any one of the many standards. Travelling around with our smartphones, tablet computers, electric toothbrushes, and so on, we would have to carry a large bag full of adaptors, and every shop selling electrical appliances would have to carry a similarly large range of adaptors.

Although that would be ridiculous, it is exactly the kind of thing we accept in the world of computing. Wikipedia lists nearly 4,000 different `extensions' for computer files, each one representing a distinct type of file.\footnote{See ``List of filename extensions'', {\em Wikipedia}, \href{http://bit.ly/28LaT4v}{bit.ly/28LaT4v}, retrieved 2016-08-16.}  A small sample is shown here:

\begin{itemize}

    \item ASE---Adobe Swatch.

    \item ART---America Online proprietary format.

    \item BLP---Blizzard Entertainment proprietary texture format.

    \item BMP---Microsoft Windows Bitmap formatted image.

    \item CD5---Chasys Draw IES image.

    \item CIT---Intergraph is a monochrome bitmap format.

    \item CPT---Corel Photo-Paint image.

\end{itemize}

Each application is severely restricted in what kinds of file it can process---it is often only one---and incompatibilities are rife, even within one area of application such as word processing or the processing of images.  A program that will run on one operating system will typically not run on any other, so normally a separate version of each program is needed for each operating system.

This kind of variety may also be found within individual files. In a Microsoft Word file, for example, there may be text in several different fonts and sizes, information generated by the ``track changes'' system, equations, WordArt, hyperlinks, bookmarks, cross-references, Clip Art, pre-defined shapes, SmartArt graphics, headers and footers, embedded Flash videos, images created by drawing tools, tables, and imported images in any of several formats including JPEG, PNG, Windows Metafile, and many more.

Variety is also alive and well amongst computer languages. Several hundred high-level programming languages are listed by Wikipedia, plus large numbers of assembly languages, machine languages, mark-up languages, style-sheet languages, query languages, modelling languages, and more.\footnote{See ``List of programming languages'', {\em Wikipedia}, \href{http://bit.ly/1GTW05W}{bit.ly/1GTW05W}, retrieved 2016-08-16; see also ``Computer language'' and links from there, {\em Wikipedia}, \href{http://bit.ly/2aZ2kag}{bit.ly/2aZ2kag}, retrieved 2016-08-17.}

Some of the variety in types of file, in formats for information within files, and in computer languages, reflects variety in the world, and is necessary and useful. But much of the variety in computing systems is quite arbitrary, without any real justification, and the source of significant problems in computing, outlined in Section \ref{problems_with_excess_variety_section}. However, despite its harmful effects, that kind of ``excess'' variety has become part of the wallpaper of computing---something that we cease to see or think about because it is so familiar, in much the same way that people once thought that applying leeches was a good way to treat an illness, or how it was accepted that the fastest way to send a letter was via a courier on horseback, with no idea that, one day, it might take only seconds to send that kind of message to the other side of the world.

\section{Problems with excess variety}\label{problems_with_excess_variety_section}

Some people may say that the variety of types of computer file, and variety in other areas of computing, is to be welcomed as a sign of vigour and creativity in the computing industry. But, often, excess variety in computing does little or nothing in terms of the user's needs or wishes, and is largely without the value of variety in art, music or literature, or in human cultures and natural languages. And excess variety in computing systems contributes to four main kinds of problem: bit rot, problems with big data, problems in the development and debugging of software, and problems with safety and cyber security.

The first of these, bit rot, is when software or data or both become unusable because technologies have moved on. Vint Cerf of Google has warned that the 21st century could become a second ``Dark Age'' because so much data is now kept in digital format, and that future generations would struggle to understand our society because technology is advancing so quickly that old files will be inaccessible.\footnote{See, for example, ``Google's Vint Cerf warns of `digital Dark Age'\thinspace'', {\em BBC News}, 2015-02-13, \href{http://bbc.in/1D3pemp}{bbc.in/1D3pemp}.}

With big data---the humongous quantities of information that now flow from industry, commerce, science, and so on---excess variety in formalisms and formats for knowledge and in how knowledge may be processed is one of several problems that make it difficult or impossible to obtain more than a small fraction of the value in those floods of data \cite{kelly_hamm_2013,national_research_council_2013}. Most kinds of processing---reasoning, pattern recognition, planning, and so on---will be more complex and less efficient than it needs to be \cite[Section III]{sp_big_data}. In particular, excess variety is likely to be a major handicap for data mining---the discovery of significant patterns and structures in big data \cite[Section IV-B]{sp_big_data}.

\sloppy Excess variety in computing also means inefficiencies in the labour-intensive and correspondingly expensive process of developing software and the difficulty of reducing or eliminating bugs in software.

And excess variety means potentially serious consequences for such things as the safety of systems that depend on computers and software, and the security of computer systems. With regard to cybersecurity, Mike Walker, head of the Cyber Grand Challenge at DARPA, has said that it counts as a grand challenge because of, {\em inter alia}, the sheer complexity of modern software.\footnote{See ``Can machines keep us safe from cyber-attack?'', {\em BBC News}, 2016-08-02, \href{http://bbc.in/2aLGwOu}{bbc.in/2aLGwOu}.}

\section{Some reforms in computing}

So what is to be done? Excess variety is a deep-rooted problem in computing as it is today and will need some radical rethinking of what computing is and how things are done. But we can catch some of the flavour of what's needed by looking at some reforms that have already been accepted and adopted in the industry.

Up until the 1970s, it was considered quite acceptable for programs to contain many of the infamous ``go to'' statements, allowing jumps from any part of a program to any other, and often leading to ``spaghetti'' programs with complex and tangled control structures that could be difficult to understand or to maintain \cite{dijkstra_1968}.\footnote{See also ``Spaghetti code'', {\em Wikipedia}, \href{http://bit.ly/1Q4AgL2}{bit.ly/1Q4AgL2}, retrieved 2016-08-03.}

Gradually, people realised that computers, like wayward children, should not be given total freedom. ``Tough love'' was needed in the shape of ``structured programming'' \cite{jackson_1975} to constrain the forms that programs could take, with benefits for comprehensibility, maintainability and reductions in cost.\footnote{See also ``Structured programming'', {\em Wikipedia}, \href{http://bit.ly/1RuSABZ}{bit.ly/1RuSABZ}, retrieved 2016-08-03.} Later, ``super-nanny'' in the shape of software gurus insisted that, for even greater benefits, computers should operate within the relative straight jacket of ``object-oriented'' programming,\footnote{Starting with {\em Simula} \cite{birtwistle_1973}, OO programming developed through {\em Smalltalk} to many other computer languages including the widely-used C++. See also ``Object-oriented programming'', {\em Wikipedia}, \href{http://bit.ly/20Rx76M}{bit.ly/20Rx76M}, retrieved 2016-08-11.} reflecting the structure of real-world things like warehouses or factories in the structure of the software that is to help manage those things.

\section{The SP theory of intelligence}\label{sp_theory_section}

These reforms have been very welcome and useful but the problem of excess variety persists. Some more tough love is needed but, fortunately, there appears to be a solution, an unexpected by-product of the {\em SP theory of intelligence} \cite{sp_extended_overview,wolff_2006} that comes with some compensating benefits including potential for the development of AI---some sugar to help the medicine go down.

The SP theory, and its realisation in the {\em SP computer model}, is the product of a long-term programme of research which has been aiming, in accordance with Occam's Razor, to simplify and integrate observations and concepts across artificial intelligence, mainstream computing, mathematics, and human perception and cognition. It is a theory of computing, a successor to Alan Turing's concept of a ``universal Turing machine'' (UTM) as a definition of ``computing'' but with much of the human-like intelligence which, as Turing recognised \cite{turing_1950,webster_2012}, is missing from the UTM.

Distinctive features and advantages of the SP theory are described in \cite{sp_alternatives}. Potential benefits and applications of the SP system are described in several papers, detailed with download links near the top of \href{http://bit.ly/1mSs5XT}{bit.ly/1mSs5XT}. It is envisaged that the SP computer model will provide the basis for a new kind of high-parallel computer, the {\em SP machine}.

In brief, the SP theory proposes: 1) that all kinds of knowledge may be represented with arrays of atomic symbols in one or two dimensions, called {\em patterns}; 2) that all kinds of processing is done by compressing information---by searching for patterns, or parts of patterns, that match each other and by the merging or ``unification'' of patterns or parts of patterns that are the same. This idea---``information compression by the matching and unification of patterns'' (ICMUP)---is bedrock in the theory; 3) more specifically, all kinds of processing is done by compressing information by the building and manipulation of {\em multiple alignments}, a concept borrowed and adapted from bioinformatics (more in Section \ref{multiple_alignments_section}); 4) Because of the intimate relationship that exists between information compression and concepts of prediction and probability \cite{li_vitanyi_2014}, the SP system is fundamentally probabilistic. That said, it has potential, if required, to imitate the clockwork style of computation in much of mathematics and logic \cite[Section 4.4.4, Chapter 10]{wolff_2006}.

As described elsewhere \cite[Section 5.2]{sp_extended_overview}, an important principle in the unsupervised learning of new knowledge in the SP system, is that it should conform to the ``DONSVIC'' principle---{\em the discovery of natural structures via information compression}, where ``natural'' structures are those that people would regard as natural, much as in object-oriented programming. There is evidence that the representation of knowledge in accordance with the DONSVIC principle normally achieves relatively high levels of information compression.

\subsection{Multiple alignments in bioinformatics and in the SP system}\label{multiple_alignments_section}

In bioinformatics, multiple alignment means an arrangement of two or more DNA sequences, or amino-acid sequences, so that, by judicious stretching of sequences, matching symbols---as many as possible---are brought into line. An example is shown in Figure \ref{DNA_figure}.

\begin{figure}[!htbp]
\fontsize{10.00pt}{12.00pt}
\centering
{\bf
\begin{BVerbatim}
  G G A     G     C A G G G A G G A     T G     G   G G A
  | | |     |     | | | | | | | | |     | |     |   | | |
  G G | G   G C C C A G G G A G G A     | G G C G   G G A
  | | |     | | | | | | | | | | | |     | |     |   | | |
A | G A C T G C C C A G G G | G G | G C T G     G A | G A
  | | |           | | | | | | | | |   |   |     |   | | |
  G G A A         | A G G G A G G A   | A G     G   G G A
  | |   |         | | | | | | | |     |   |     |   | | |
  G G C A         C A G G G A G G     C   G     G   G G A
\end{BVerbatim}
}
\caption{A `good' multiple alignment amongst five DNA sequences.}
\label{DNA_figure}
\end{figure}

\begin{figure}[!htbp]
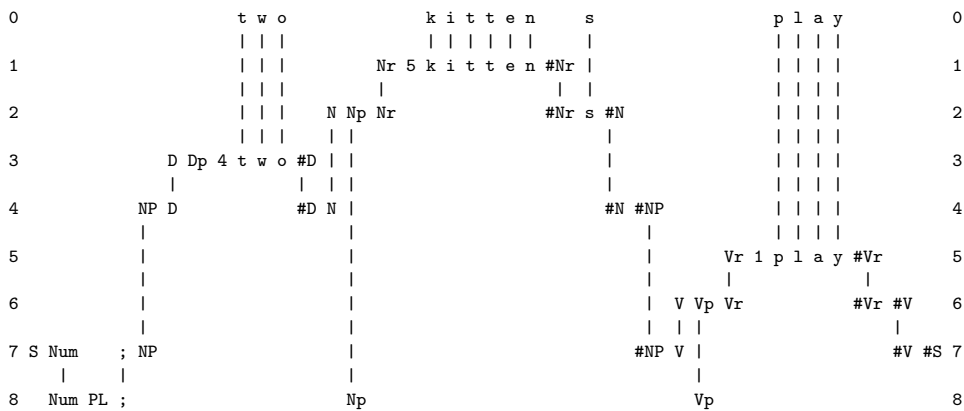

\fontsize{07.50pt}{09.00pt}
\centering
{\bf
\begin{BVerbatim}
0                      t w o              k i t t e n     s                  p l a y           0
                       | | |              | | | | | |     |                  | | | |
1                      | | |         Nr 5 k i t t e n #Nr |                  | | | |           1
                       | | |         |                 |  |                  | | | |
2                      | | |    N Np Nr               #Nr s #N               | | | |           2
                       | | |    | |                         |                | | | |
3               D Dp 4 t w o #D | |                         |                | | | |           3
                |            |  | |                         |                | | | |
4            NP D            #D N |                         #N #NP           | | | |           4
             |                    |                             |            | | | |
5            |                    |                             |       Vr 1 p l a y #Vr       5
             |                    |                             |       |             |
6            |                    |                             |  V Vp Vr           #Vr #V    6
             |                    |                             |  | |                   |
7 S Num    ; NP                   |                            #NP V |                   #V #S 7
     |     |                      |                                  |
8   Num PL ;                      Np                                 Vp                        8
\end{BVerbatim}
}
\caption{The best multiple alignment created by the SP computer model with the sentence `t w o k i t t e n s p l a y' as the New pattern and a set of Old patterns representing grammatical structures, including words.}
\label{alignments_figure_1a}
\end{figure}

In the SP system, the multiple alignment concept has been adapted as illustrated in Figure \ref{alignments_figure_1a}. Here, the pattern in row 0---a simple sentence in this example---is input from the system's environment and is classified as ``New'' information. The patterns in rows 1 to 8---which, in this example, represent grammatical structures including words---are part of a relatively large set of stored patterns which are classified as ``Old''.

The aim is to find a multiple alignment, or sometimes more than one, that provides a means of encoding the New information economically in terms of Old patterns (\cite[Section 4.1]{sp_extended_overview}, \cite[Section 3.5]{wolff_2006}). In this example, the best multiple alignment, shown in the figure, may be seen as an analysis or parsing of the input sentence in terms of the stored grammatical structures.

The multiple alignment concept, as it has been developed in the SP programme of research, has proved to be remarkably versatile in the representation of diverse forms of knowledge and modelling diverse aspects of intelligence, as outlined below. It has potential to be the ``double helix'' of intelligence---as significant for an understanding of ``intelligence'' broadly construed as is DNA for biological sciences.

\section{How the SP system may help solve the problem of excess variety in computing}

There are three main reasons why the SP theory may help solve the problem of excess variety in computing: the versatility of the SP system in the representation of diverse kinds of knowledge; the versatility of the system in modelling diverse aspects of intelligence; and the generality of the principles on which the SP system is based. Versatility and generality in the system can yield a global simplification in computing, as described in Section \ref{how_to_achieve_simplification_section}.

\subsection{Versatility of the SP system in the representation of knowledge}\label{versatility_in_kr_section}

\sloppy SP patterns, within the multiple alignment framework, have proved to be an effective means of representing several different kinds of knowledge, including the syntax of natural languages, class hierarchies, part-whole hierarchies, discrimination networks and trees, entity-relationship structures, relational knowledge, rules for reasoning, patterns, images, structures in three dimensions, and procedural knowledge. There is more detail throughout \cite{wolff_2006} and \cite{sp_extended_overview}, and there are references to further sources of information in \cite[Section III-B]{sp_big_data}.

\subsection{Versatility of the SP system in aspects of intelligence}\label{versatility_in_intelligence_section}

The processing of knowledge in the multiple alignment framework has proved to be a means of modelling several aspects of intelligence including unsupervised learning, the processing of natural language, fuzzy pattern recognition, recognition at multiple levels of abstraction, best-match and semantic forms of information retrieval, planning, problem solving, and several kinds of reasoning including: one-step `deductive' reasoning, chains of reasoning, abductive reasoning, reasoning with probabilistic networks and trees, reasoning with `rules', nonmonotonic reasoning, Bayesian reasoning with ``explaining away'', causal reasoning, and reasoning that is not supported by evidence (\cite[Chapters 5 to 9]{wolff_2006}, \cite[Section 10]{sp_extended_overview}). The system also has potential in inference via inheritance of attributes (\cite[Section 9.2]{sp_extended_overview}, \cite[Section 6.4]{wolff_2006}), spatial reasoning \cite[Section IV-F.1]{sp_autonomous_robots}, and what-if reasoning \cite[Section IV-F.2]{sp_autonomous_robots}.

\subsection{Generality of the SP system}\label{generality_of_sp_section}


There is reason to believe that, in addition to its strengths in the representation of knowledge and in aspects of artificial intelligence, the SP system may be a vehicle for any kind of computing:

\begin{itemize}

    \item {\em The generality of ICMUP and information compression via multiple alignment}. That the SP system should have wide scope is suggested by:

        \begin{itemize}

            \item The generality of information compression by the matching and unification of patterns (ICMUP), and, more specifically, information compression via multiple alignment, in the representation of knowledge and, more specifically, in the realisation of the DONSVIC principle (Section \ref{sp_theory_section}).

            \item The significance of information compression in its intimate connection with concepts of prediction and probability \cite{li_vitanyi_2014}, mentioned in Section \ref{sp_theory_section}.

        \end{itemize}

    \item {\em Turing completeness}. As described in \cite[Chapter 4]{wolff_2006}, the workings of the SP system may be interpreted in terms of the operations of a Post canonical system \cite{post_1943}. Since it is accepted that the Post canonical system is Turing complete \cite[Chapters 10 to 14]{minsky_1967}---meaning that it can simulate any any single-taped universal Turing machine---the same is probably true of the SP system.

    \item {\em Modelling programming concepts in the multiple alignment framework}. Although multiple alignments like the one shown in Figure \ref{alignments_figure_1a} may seem to be far removed from the programming of ordinary computers, the relationship is much closer than it may superficially appear. The SP system with the multiple alignment framework can not only model ``static'' kinds of knowledge structure like class hierarchies and part-whole hierarchies (Section \ref{versatility_in_kr_section})
        but, as described in \cite[Section 6.6]{sp_benefits_apps}, it can also model most of the concepts that are familiar in ordinary programming, including {\em procedure}, {\em variable}, {\em value}, {\em type}, {\em function with parameters}, {\em conditional statement}, {\em iteration or recursion}, and the elements of {\em object-oriented programming}. There is also potential for the processing of parallel streams of information as described in \cite[Sections V-G, V-H, and V-I, and Appendix C]{sp_autonomous_robots}.

    \item {\em Compression of information}. Although the SP system may, in principle, be a vehicle for any kind of computing---because information compression is a broad church that may encompass low levels of information compression as well as high levels---it would normally operate in a manner that would give priority to well-compressed structures, avoiding the poorly-compressed structures which are prominent in computing systems as they are now, with excess variety and complexity, and the associated problems outlined in Section \ref{problems_with_excess_variety_section}.

\end{itemize}

With regard to the third point---the modelling of programming concepts---there is the possibility that, with further development, the workings of the SP system would be determined largely by what it learns via unsupervised learning (an important feature of the SP system), but also, where necessary, by a version of ``programming'' that would be similar in some respects to how computers are programmed now.

Key differences between SP programming and traditional programming would be:

 \begin{itemize}

    \item {\em Real-world structures and procedures}. That the former should be concerned exclusively with structures and procedures in the world outside the computer---such as the control of traffic lights or processes in a factory---whereas the latter is concerned partly with aspects of the world outside the computer and partly with overcoming deficiencies in computing hardware. We shall return to the latter point in Section \ref{how_to_achieve_simplification_section}.

    \item {\em Parsimony in the representation of knowledge}. That the former, in accordance with the principles of object-oriented programming and the DONSVIC principle (Section \ref{sp_theory_section}), should aim to model real-world structures and processes in an economical manner, whereas the latter, in its concern with the workings of computing hardware, may lose touch with the need for parsimony in the representation of knowledge.

 \end{itemize}


\subsection{Towards a universal framework for the representation and processing of diverse kinds of knowledge}

Overall, the three aspects of generality in the SP system outlined above suggest that it's potential is not restricted to the areas mentioned in
Sections \ref{versatility_in_kr_section} and \ref{versatility_in_intelligence_section}
but may extend to all kinds of knowledge and all aspects of computation and human-like intelligence. It has potential to be a {\em universal framework for the representation and processing of diverse kinds of knowledge} (UFK), as outlined in \cite[Section III-A]{sp_big_data}.

As a UFK, the SP system would be ``universal'' in the sense that it would provide for the economical representation of any kind of knowledge and for any kind of computation, including the kinds of things that are seen as human-like intelligence---but it would provide more discipline than in present-day computers, reducing or eliminating unnecessary complexity in computing and the kinds of excess variety discussed earlier. It has the potential to reduce the near-4000 different kinds of computer file to one, and to reduce the hundreds of different computer languages to one.

An analogy is that, with clay, we can in principle create any shape. But experts in the creation of ceramics know that some kinds of shape work better than others. Likewise, in computing, we should be seeking to avoid the excessive complexity that features in so much of modern software.

\section{How a global simplification in computing may be achieved}\label{how_to_achieve_simplification_section}

Another way of looking at these issues is via an idea that is already established in computer science as the basis for such things as database management systems (DBMSs) and shells for expert systems. A lot of effort can be saved with DBMSs and a lot of complexity can be avoided by creating one general-purpose system for the storage and retrieval of data and loading it with different kinds of data according to need. Effort can be saved because there is no need, with each new database, to re-program the procedures for the storage and retrieval of data. And, correspondingly, there will be an overall reduction in the complexity of any collection of several DBMSs. Much the same may be said, {\em mutatis mutandis}, about expert systems.

The SP system is more ambitious than DBMSs and shells for expert systems because, instead of the fairly narrow range of capabilities of those systems, the SP system aims to provide a much wider variety of capabilities, with human-like versatility and adaptability in intelligence and, where required, the means of modelling real-world procedures and processes in the manner of ordinary programming (Section \ref{generality_of_sp_section}).

\begin{figure}[!htbp]
\centering
\includegraphics[width=0.9\textwidth]{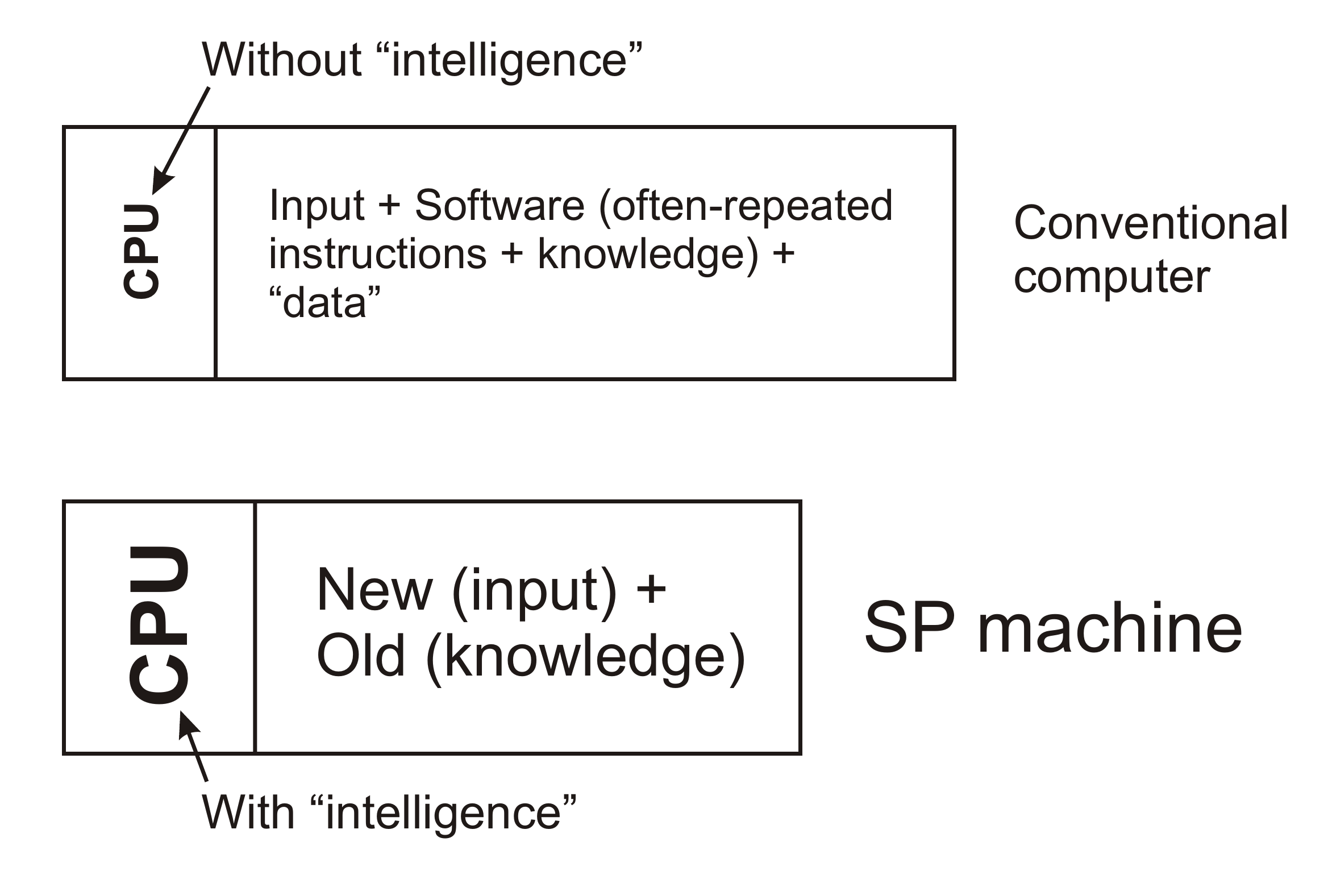}
\caption{Schematic representations of a conventional computer and the proposed {\em SP machine}, showing potential benefits in terms of simplification, as discussed in the text. Adapted from Figure 4.7 in \protect\cite{wolff_2006}, with permission.}
\label{two_schematic_computers_figure}
\end{figure}

Potential benefits in terms of simplicity are illustrated in Figure \ref{two_schematic_computers_figure} which shows, at the top, a schematic representation of a conventional computer and, at the bottom, a schematic representation of the SP machine, expressing the SP theory.

In the conventional computer there is a central processing unit (CPU), with little or no human-like intelligence---shown on the left in the figure. On the right, there is ``input'' to the computer---a query or some similar smallish piece of information to be processed; there is also ``software'' with elements described in the next paragraph; and, very often, there is the kind of ``data'' that may stored in an external file or database.

The software normally contains two main kinds of information:

\begin{itemize}

    \item Instructions that are designed to make up for the deficiencies in the CPU. These may include procedures for recognising patterns, procedures for searching for information, and procedures for retrieving stored information. Very often, such procedures, or variants of them, are repeated again and again within one program or across many different programs.

    \item Very often, the software also contains significant knowledge about the world, such as real-world procedures for applying for a driving licence, booking a seat on a train, and so on.

\end{itemize}

It is envisaged that the SP machine will be simpler. All processing will be done in a CPU, shown on the left in the figure, which will supply all the human-like intelligence of the system, including procedures for the building and manipulation of multiple alignments. The rest of the system, shown on the right, will be New and Old information as described in Section \ref{multiple_alignments_section}.

The New information is ``input'' to the system, as in the conventional computer, while the Old information is the system's knowledge about the world, including what would conventionally be called ``data'' and the kinds of knowledge mentioned earlier such as real-world procedures for applying for a driving licence or for booking a seat on a train.\footnote{With applications that are purely abstract, such as applications that are specialised for one or more aspects of mathematics, we may suppose that the Old information would comprise knowledge of relevant abstract concepts.}

The key differences between a conventional computer and the SP machine are that, in the latter, all information about how to process information would be contained in the CPU and there would be a single repository of knowledge about the world, including knowledge about real-world procedures and processes.

The CPU in the SP machine is shown a bit larger than the CPU in the conventional computer to suggest that it is a bit more complex than a conventional CPU. But, despite this additional complexity, the rectangle representing the SP machine is shown smaller than the rectangle representing the conventional computer to indicate that, with the SP system, there can be a global simplification in computing. This is because it will not be necessary to add instructions, often with repetition, to make up for the shortcomings of the CPU in the conventional computer, and also because all knowledge in the SP machine will be highly compressed.

Readers may say ``Isn't this simply a reinvention of the concept of declarative programming, separating information about `what' a program is to do from the details of `how' the objectives are to be achieved?''\footnote{See ``Declarative programming'', {\em Wikipedia}, \href{http://bit.ly/2aVJAIE}{bit.ly/2aVJAIE}, retrieved 2016-08-15.} Yes and no. ``Yes'' because there are some similarities in the overall concept but ``no'' because the SP machine, with multiple alignment at its core, has the potential to yield much more human-like versatility and adaptability than logic programming, functional programming, or the like.

\section{Conclusion}

The SP system was developed mainly to advance AI. But in addition to its several strengths in that area, it has the potential to reduce or eliminate the curse of variety in computing. Although it may seem impossibly ambitious, {\em there is real potential to cut the variety of file types from nearly 4,000 to one, and to cut the hundreds of computer languages to one}.

The multiple alignment framework has the potential to be a {\em universal framework for the representation and processing of diverse kinds of knowledge} (UFK). It is envisaged that all kinds of knowledge will be represented with SP patterns and all kinds of processing will be done via the building and manipulation of multiple alignments.

Of course, there would still be different kinds of application. But instead of programs containing a mixture of real-world knowledge and often-repeated instructions needed to make good the shortcomings of conventional CPUs, each application will comprise nothing but knowledge that is relevant to its area of application, including knowledge about significant real-world entities, and classes of such entities, and real-world operations with those things.

\begin{figure}[!htbp]
\centering
\includegraphics[width=0.9\textwidth]{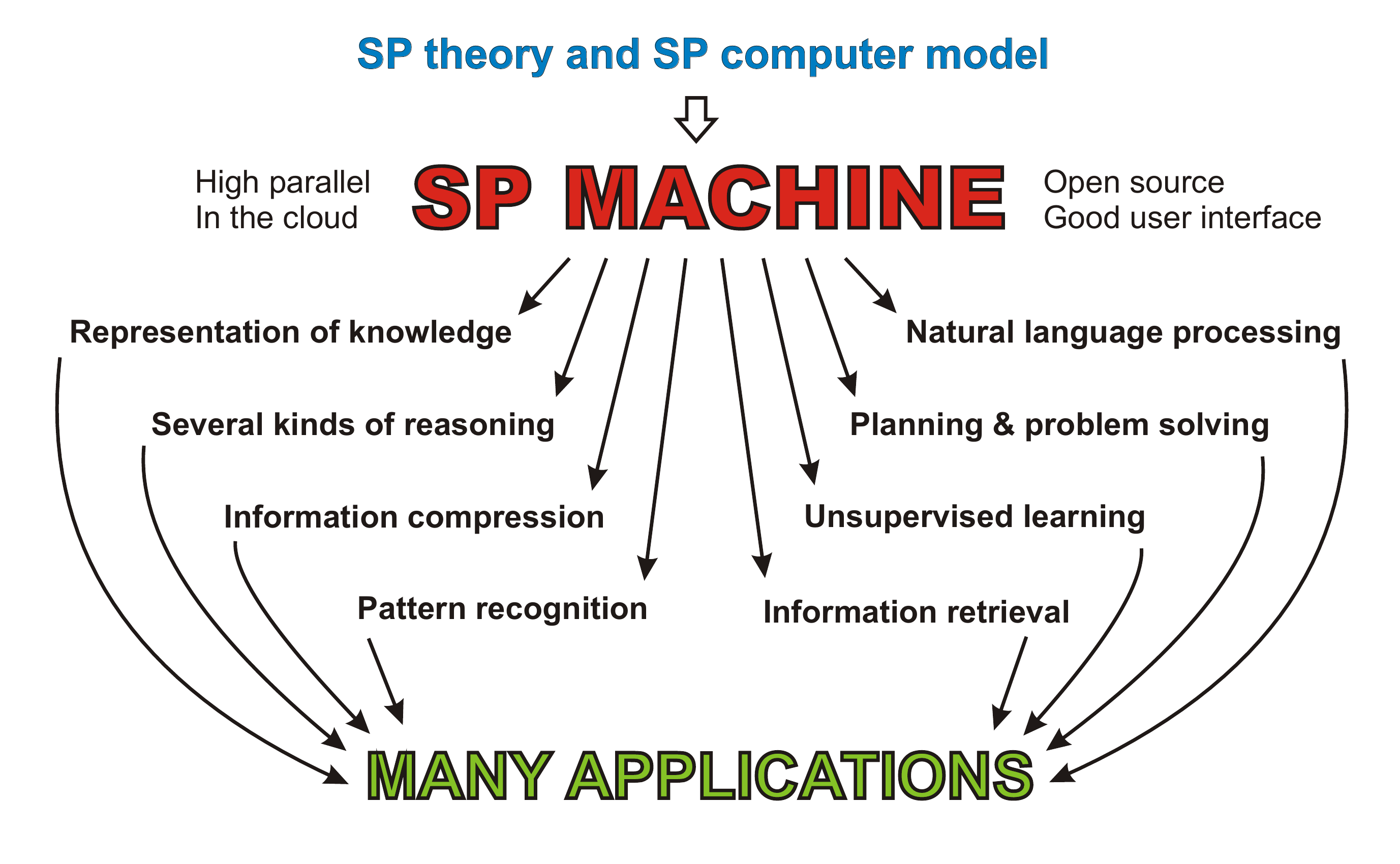}
\caption{A schematic view of how the SP machine may develop from the SP theory and the SP computer model. Adapted, with permission, from Figure 6 in \protect\cite{sp_big_data}.}
\label{sp_machine_development_figure}
\end{figure}

Probably, the best way to advance these ideas would be, firstly, to create a high-parallel version of the SP machine, based on the SP theory as it has been realised in the SP computer model, and, secondly, to make the SP machine available to researchers everywhere to see what can be done with the SP system, and to create new versions of it. How things may develop is shown schematically in Figure \ref{sp_machine_development_figure}.

\section{Author}

Gerry Wolff (jgw@cognitionreseardh.org) is Director of CognitionResearch.org (\href{http://www.cognitionresearch.org/}{www.cognitionresearch.org}), concentrating mainly on the development of the {\em SP theory of intelligence} and its realisation in the {\em SP machine}.

\bibliographystyle{plain}

\end{document}